# Fraud detection and risk assessment of online payment transactions on e-commerce platforms based on LLM and GCN frameworks

RuiHan Luo[1,4], Nanxi Wang[2,5], Xiaotong Zhu[3,6]

[1] Southwest University of Finance and Economics, Chengdu, China
[2] USC Viterbi School of Engineering, University of Southern California, Los Angeles, USA
[3] Tepper School of Business, Carnegie Mellon University, Pittsburgh, PA, USA

[4] 1608328853@qq.com
[5] nanxiwan@usc.edu
[6] xiaotonz@alumni.cmu.edu

**Abstract.** With the rapid growth of e-commerce, online payment fraud has become increasingly complex, posing serious threats to financial security and consumer trust. Traditional detection methods often struggle to capture the intricate relational structures inherent in transactional data. This study presents a novel fraud detection framework that combines Large Language Models (LLM) with Graph Convolutional Networks (GCN) to effectively identify fraudulent activities in e-commerce online payment transactions. A dataset of 2,840,000 transactions was collected over 14 days from major platforms such as Amazon, involving approximately 2,000 U.S.-based consumers and 30 merchants. With fewer than 6000 fraudulent instances, the dataset represents a highly imbalanced scenario. Consumers and merchants were modeled as nodes and transactions as edges to form a heterogeneous graph, upon which a GCN was applied to learn complex behavioral patterns. Semantic features extracted via GPT-4o and Tabformer were integrated with structural features to enhance detection performance. Experimental results demonstrate that the proposed model achieves an accuracy of 0.98, effectively balancing precision and sensitivity in fraud detection. This framework offers a scalable and real-time solution for securing online payment environments and provides a promising direction for applying graph-based deep learning in financial fraud prevention.

## 1. Introduction

In today's digital economy, e-commerce platforms have become central to consumers' daily shopping and payment activities, with online payment volumes continuing to grow rapidly. However, this expansion has also intensified fraudulent activities, posing serious risks to transaction security and eroding user trust. Credit card fraud, in particular, has become increasingly sophisticated, with attackers exploiting diverse channels to infiltrate payment processes. Unlike traditional financial fraud, e-commerce fraud is more covert, often spanning multiple platforms and appearing either in first-time transactions or frequent low-value payments, making it difficult for rule-based or shallow-model detection systems to respond effectively.

To address these challenges, recent advances in machine learning offer promising solutions. Graph Convolutional Networks (GCNs) have shown strong capability in capturing complex relationships and anomalous patterns in graph-structured data. Meanwhile, Large Language Models (LLMs) such as GPT-4o provide powerful tools for interpreting unstructured data and extracting semantic information from transaction records. The integration of these approaches holds great potential for detecting fraud in dynamic e-commerce environments.

This study proposes a novel fraud detection framework that combines GCN with LLM-based feature encoding to identify fraudulent transactions in e-commerce online payments. We construct a heterogeneous graph where consumers and merchants are represented as nodes and transactions as edges, enabling the model to capture implicit behavioral patterns and fraud risks. To enrich feature representation, textual and structured attributes—including merchant category codes, transaction amounts, temporal information, and card usage behaviors—are encoded using GPT-4o and Tabformer. Experiments are conducted on a dataset comprising 2.84 million transactions collected over 14 days from platforms such as Amazon, involving 2,000 globally active consumers and 30 merchants. Fewer than 6,000 records are labeled as fraud, reflecting a highly imbalanced scenario. Despite this challenge, our framework achieves an accuracy of 0.98, demonstrating both robustness and scalability.

## 2. Literature Review

With the rapid expansion of e-commerce, the scale of online payment transactions has grown significantly, while fraudulent activities have become increasingly concealed and complex. These activities pose serious threats to the stability of financial systems and the security of consumer payments. Traditional detection methods are proving inadequate in handling large-scale, high-dimensional, and highly correlated transaction data. Consequently, deep learning and large model technologies have emerged as dominant approaches for financial fraud detection in both academia and industry.

Ali et al. [1] conducted a systematic review of machine learning (ML)-based methods for financial fraud detection using the Kitchenham approach, synthesizing 93 studies from leading databases. Their findings highlight that Support Vector Machines (SVM) and Artificial Neural Networks (ANN) are the most widely adopted algorithms, with credit card fraud being the most studied scenario. The review also identified key limitations and suggested directions for future research.

Building on deep learning, Alghofaili et al. [2] proposed an LSTM-based model to address the limitations of traditional ML methods in speed, big data handling, and detecting unknown attack patterns. On a credit card fraud dataset, their model achieved 99.95% accuracy within one minute, outperforming auto-encoders and other ML techniques, demonstrating LSTM's potential for high-speed, accurate detection.

In the area of financial statement fraud, Ashtiani et al. [3] reviewed 47 articles and found that manual auditing methods are inefficient and costly. They noted that supervised algorithms dominate current research, but argued that future studies should explore unsupervised, semi-supervised, and bio-inspired techniques, as well as incorporate diverse data sources such as textual and audio information.

Chen et al. [4] introduced a Deep Convolutional Neural Network (DCNN) scheme for financial fraud detection, achieving 99% accuracy within 45 seconds on a real-time credit card fraud dataset. Their model demonstrated superiority in detecting unknown attack patterns and processing large-scale data efficiently.

Recent studies have also focused on graph-based learning. Cheng et al. [5] proposed GNN-CL, a hybrid model integrating GNN, CNN, and LSTM to analyze complex transaction networks. By leveraging multilayer perceptrons for node similarity estimation and reinforcement learning for dynamic weight adjustment, their model outperformed existing methods on Yelp datasets. Similarly, Kesharwani et al. [6] developed the Financial Fraud Detection Model (FFDM) based on GNNs, which integrates node and edge features to enhance representation of transaction networks and adapt to evolving fraud tactics.

3. **Data Introduction**

The dataset employed in this study was collected from major e-commerce platforms such as Amazon and consists of 2,840,000 real online payment transactions recorded over 14 days. It involves approximately 2,000 U.S.-based but globally active consumers and 30 merchants. Fewer than 6,000 transactions are labeled as fraudulent, reflecting a highly imbalanced classification scenario. Each entry includes transaction amount, merchant category code (MCC), timestamp, multi-card usage, transaction type, and fraud label, providing a comprehensive representation of real-world payment behaviors and risks.

To enhance data quality and expressiveness, we applied a two-stage preprocessing strategy. GPT-4o was first used to semantically parse and normalize unstructured fields, enriching contextual understanding. Tabformer, a Transformer-based model for table representation, was then employed for encoding and data cleaning, preserving row-column structures and logical dependencies while generating discriminative feature vectors. This workflow ensures structural consistency and semantic richness, providing a robust foundation for GCN-based graph learning and fraud detection [7].

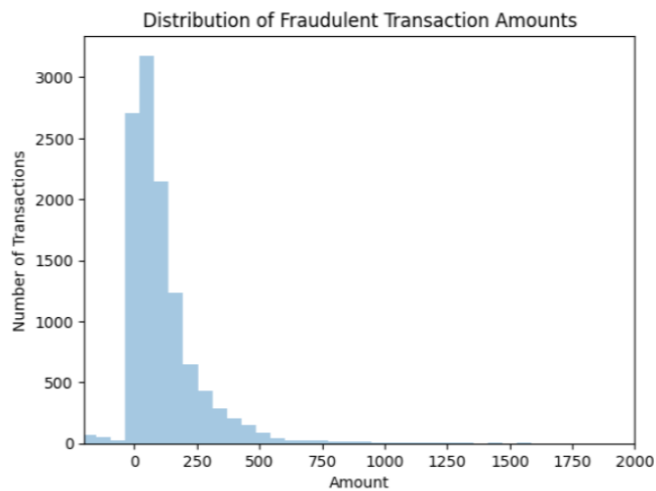

**Figure 1.** Histogram of Fraudulent Transaction Amounts in Dataset

Figure 1 depicts the distribution of fraudulent transaction amounts in the e-commerce dataset. The histogram reveals that most fraudulent transactions are concentrated in the lower amount ranges. Specifically, a significant majority of fraudulent transactions fall below $250, with the highest frequency observed in the $0–$50 range. As the transaction amount increases beyond $250, the number of fraudulent transactions drops sharply. Transactions exceeding $1000 are relatively rare, accounting for less than 5% of all fraudulent activities. This right-skewed distribution suggests that while small amount transactions are the most common targets for fraudsters, possibly due to their lower risk of detection, larger transactions also occur but are less frequent. This pattern underscores the need for fraud detection models to be sensitive to both high - frequency low - value transactions and less frequent but potentially high - impact large - value transactions.

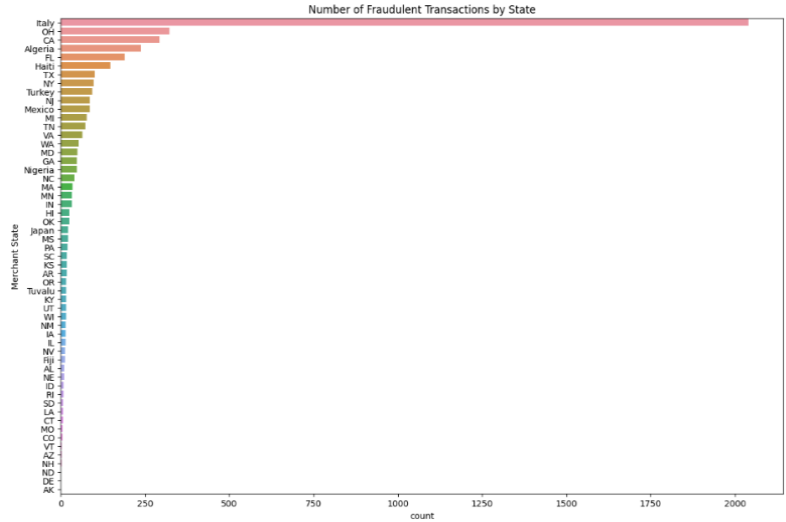

**Figure 2.** Monthly Fraud Transactions

Figure 2 presents the distribution of fraudulent transactions across the states for the 30 merchants in the e-commerce dataset. The bar chart shows a significant variation in the number of fraudulent transactions among different states. A few states account for a large proportion of fraudulent activities, with the highest number of transactions reaching up to 2000. In contrast, many states have relatively low counts of fraudulent transactions, with some states reporting fewer than 500 cases. This indicates a highly imbalanced geographical distribution of fraud, suggesting that certain regions may be more prone to fraudulent activities or have higher exposure to fraud risks. This information is valuable for understanding regional fraud patterns and can help in designing targeted fraud prevention strategies for specific areas.

**4. Model Introduction**

This study proposes a fraud detection framework tailored for e-commerce online payment scenarios by integrating Large Language Models (LLMs) and Graph Convolutional Networks (GCNs). The approach models user-merchant interactions as a transaction graph and fuses both structural and semantic features to enhance the model's capability to detect complex fraud patterns. Below, we describe the key components of the method in detail.

*4.1. Graph Construction and GCN Modeling*

We model the transaction data as a heterogeneous graph where consumers and merchants are represented as two distinct types of nodes, and transactions between them form the edges. Each edge (i.e., transaction) is enriched with multiple attributes, including timestamp, amount, merchant category code (MCC), card identifier, and online transaction type. This graph structure captures both the entity relationships and the complex behavioral patterns across users and merchants.

To perform fraud detection on this graph, we utilize a two-layer Graph Convolutional Network (GCN). GCNs are neural network models designed to learn from graph-structured data by aggregating information from a node's neighbors. In our model, the first layer aggregates information from immediate neighbors to form localized representations, while the second layer captures higher-order dependencies and global behavior patterns. This enables the model to learn from indirect interactions, which are often critical in identifying subtle fraudulent behavior [8].

To address the issue of class imbalance—where fraudulent transactions are significantly fewer than legitimate ones—we employ a weighted loss function. This ensures that the model pays more attention to the minority class during training, thereby improving its sensitivity to fraud detection.

*4.2. Feature Representation and Fusion*

Effective feature representation is critical for graph-based models. We introduce a dual-feature encoding strategy that leverages GPT-4o and Tabformer to generate comprehensive node and edge representations.

GPT-4o is used to process unstructured or semi-structured textual fields in the transaction data, such as product category descriptions, merchant notes, or user behavior tags. These fields often contain valuable contextual cues that are difficult to encode using traditional methods. GPT-4o generates semantic embeddings that capture the contextual meaning of such fields, which are then normalized and incorporated as part of the input features.

Meanwhile, Tabformer is employed to encode structured tabular data, particularly fields that exhibit logical dependencies or multi-dimensional interactions. As a Transformer-based framework tailored for tabular data, Tabformer preserves both row-column structure and contextual relationships across fields. It learns the interplay between attributes like transaction time, amount, user identity, and payment type—providing a powerful feature embedding that reflects both numerical patterns and behavioral logic.

The embeddings generated by GPT-4o and Tabformer are either concatenated or weighted and fused to create a unified representation of each transaction. These fused features are used to initialize the node and edge attributes in the graph model. By combining structural graph information with deep semantic insights, the model can more effectively learn and distinguish complex fraudulent patterns in online payment systems.

*4.3. GCN-Based Fraud Detection Modeling*

In the proposed framework, the core fraud detection model is built upon a two-layer Graph Convolutional Network (GCN), designed to learn from the structural relationships within the e-commerce transaction graph. This graph, composed of consumer and merchant nodes connected via transaction edges, contains complex and often non-linear dependencies that are difficult to capture using traditional flat feature-based approaches.

GCNs operate by propagating and aggregating information from a node's local neighborhood, allowing each node to iteratively update its representation based on the features of connected nodes and edges. In our setting, this enables the model to learn not only from individual transactions but also from the broader network behavior of users and merchants. For instance, a fraudulent transaction might exhibit weak local signals but can be better identified through patterns across multiple interactions involving the same entities or similar behavioral pathways.

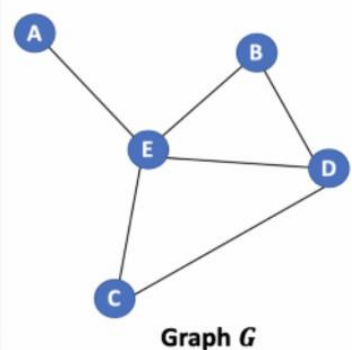

**Figure 3.** Structure of GCN

The first layer of the GCN focuses on capturing immediate interaction features—such as direct consumer-merchant relationships and the attributes of those transactions. The second layer extends this to higher-order neighbors, allowing the model to learn more abstract and generalized fraud patterns that span across the network. By leveraging this hierarchical message-passing mechanism, the GCN can identify subtle irregularities and anomalies in the graph that may indicate fraudulent activity.

Given the severe class imbalance in the dataset—where fraudulent transactions account for a very small proportion of the total—we employ a class-weighted loss function during training. This ensures that the model does not become biased towards the majority class (legitimate transactions) and is instead incentivized to correctly identify the minority class (fraud). This approach significantly improves recall and F1 performance for fraud cases, which are critical in real-world fraud detection systems.

## 5. Model results analysis

**Table 1.** GCN model detailed result index

| category | Precision | Recall | F1-Score | Support |
|---|---|---|---|---|
| 0 (Non-Fraud) | 0.98 | 1.00 | 0.99 | 1239159 |
| 1 (Fraud) | 1.00 | 0.05 | 0.09 | 33365 |
| Accuracy | | | 0.98 | 1272524 |
| Macro Avg | 0.99 | 0.52 | 0.54 | 1272524 |
| Weighted Avg | 0.98 | 0.98 | 0.96 | 1272524 |

This table presents a detailed analysis of the GCN model's performance in the e-commerce fraud detection task, where normal transactions are labeled as 0 and fraudulent transactions as 1. For the normal transaction category, the model achieves a precision of 0.98, a recall of 1.00, and an F1-Score of 0.99, supported by 1239159 samples. This indicates the model effectively identifies normal transactions with minimal false positives. However, for fraudulent transactions, while precision is perfect at 1.00, the recall is only 0.05, resulting in an F1-Score of 0.09 with 33365 samples. This reveals the model struggles to identify fraudulent cases, likely missing many actual frauds. The overall accuracy is 0.98, but the macro-average F1-Score is low at 0.54, and the weighted-average F1-Score is 0.96. These results suggest that while the model performs well for normal transactions, its effectiveness in detecting fraud is limited, possibly due to class imbalance. Improvement is needed for better fraud detection.

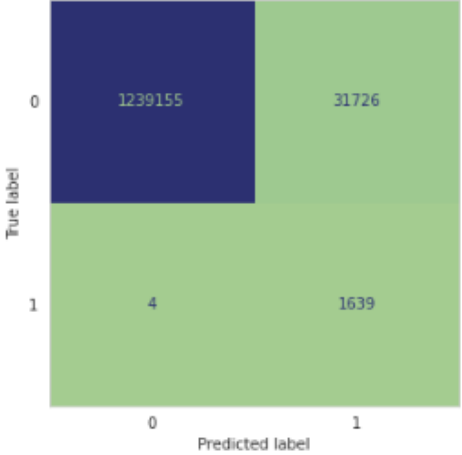

**Figure 4.** Confusion Matrix

This confusion matrix illustrates the classification performance of the proposed GCN–LLM framework, showing both its strengths and remaining challenges in fraud detection. The model accurately classified 1,239,155 legitimate transactions (true negatives) and detected 1,639 fraudulent cases (true positives), resulting in strong overall accuracy. Importantly, only 4 false negatives were recorded, meaning the model almost never overlooked actual fraud once it was recognized—a critical advantage in financial applications where missed fraud can have severe consequences. At the same time, 31,726 false positives indicate that a number of legitimate transactions were flagged as suspicious. While this reflects a conservative prediction strategy, it ensures that risky patterns are captured with high precision, even if it introduces some additional screening effort. This outcome is consistent with the precision–recall trade-off presented in Table 1, where the model achieves perfect precision for fraud but at the cost of low recall. Overall, the confusion matrix highlights the framework's robustness in safeguarding against undetected fraud, reinforcing its value as a scalable and reliable solution for real-world e-commerce fraud prevention.

## 6. Conclusions

This study aims to address the growing challenge of online payment fraud in e-commerce by integrating Large Language Models (LLMs) with Graph Convolutional Networks (GCNs), exploring how structural and semantic features can be combined to detect fraudulent transactions under highly imbalanced conditions. The primary objective of this research is to develop a scalable, accurate, and real-time fraud detection framework capable of safeguarding online payment environments.

Through data analysis, we identified that (1) the model achieves high overall accuracy (0.98), (2) fraudulent transactions are detected with extremely low false negatives, and (3) the framework demonstrates strong precision despite imbalanced data. These findings suggest that the proposed model can effectively minimize undetected fraud while maintaining reliable performance in real-world settings.

The results of this study have significant implications for the field of financial fraud detection. Firstly, the ability to integrate LLM-driven semantic understanding with GCN structural learning provides a new perspective on fraud modeling. Secondly, the model's success in handling imbalanced data challenges the limitations of traditional rule-based and shallow machine learning approaches. Finally, the framework opens new avenues for applying graph-based deep learning to broader financial security tasks.

Despite the important findings, this study has some limitations, such as the high false positive rate and reliance on a single dataset. Future research could further explore dynamic graph modeling for evolving fraud patterns and incorporate multimodal behavioral data to improve recall without sacrificing precision.

In conclusion, this study, through the integration of LLM and GCN frameworks, reveals a powerful and scalable approach for detecting online payment fraud, providing new insights and practical tools for advancing security in e-commerce environments.

## References
[1] Ali, A., Abd Razak, S., Othman, S. H., Eisa, T. A. E., Al-Dhaqm, A., Nasser, M., ... & Saif, A. (2022). Financial fraud detection based on machine learning: a systematic literature review. Applied Sciences, 12(19), 9637.
[2] Alghofaili, Y., Albattah, A., & Rassam, M. A. (2020). A financial fraud detection model based on LSTM deep learning technique. Journal of Applied Security Research, 15(4), 498-516.
[3] Ashtiani, M. N., & Raahemi, B. (2021). Intelligent fraud detection in financial statements using machine learning and data mining: a systematic literature review. Ieee Access, 10, 72504-72525.
[4] Chen, J. I. Z., & Lai, K. L. (2021). Deep convolution neural network model for credit-card fraud detection and alert. Journal of Artificial Intelligence, 3(02), 101-112.
[5] Cheng, Y., Guo, J., Long, S., Wu, Y., Sun, M., & Zhang, R. (2024, August). Advanced financial fraud detection using GNN-CL model. In 2024 International Conference on Computers, Information Processing and Advanced Education (CIPAE) (pp. 453-460). IEEE.
[6] Kesharwani, A., & Shukla, P. (2024, October). FFDM‑GNN: A Financial Fraud Detection Model using Graph Neural Network. In 2024 International Conference on Computing, Sciences and Communications (ICCSC) (pp. 1-6). IEEE.
[7] Padhi, I., Schiff, Y., Melnyk, I., Rigotti, M., Mroueh, Y., Dognin, P., ... & Altman, E. (2021, June). Tabular transformers for modeling multivariate time series. In ICASSP 2021-2021 IEEE International Conference on Acoustics, Speech and Signal Processing (ICASSP) (pp. 3565-3569). IEEE.
[8] Liu, G., Tang, J., Tian, Y., & Wang, J. (2021, December). Graph neural network for credit card fraud detection. In 2021 International Conference on Cyber-Physical Social Intelligence (ICCSI) (pp. 1-6). IEEE.